\documentclass[12pt]{iopart}
\usepackage{graphicx}

\begin{document}


\title[Transport in single-molecule transistors]{Transport in single-molecule transistors:  Kondo physics
and negative differential resistance}

\author{Lam H. Yu and Douglas Natelson}

\address{Department of Physics and Astronomy, Rice University, 6100 Main St., Houston, TX 77005}

\date{\today}

\begin{abstract}
We report two examples of transport phenomena based on sharp features
in the effective density of states of molecular-scale transistors:
Kondo physics in C$_{60}$-based devices, and gate-modulated negative
differential resistance (NDR) in ``control'' devices that we ascribe
to adsorbed contamination.  We discuss the need for a statistical
approach to device characterization, and the criteria that must be
satisfied to infer that transport is based on single molecules.  We
describe apparent Kondo physics in C$_{60}$-based single-molecule
transistors (SMTs), including signatures of molecular vibrations in
the Kondo regime.  Finally, we report gate-modulated NDR in 
devices made without intentional molecular components, and discuss
possible origins of this property.
\end{abstract}

\ead{natelson@rice.edu}

\maketitle


\section{Introduction}

Though active electronic devices made from molecules were originally
suggested thirty years ago\cite{AvirametAl74CPL}, interfacing small
numbers of molecules with metal electrodes for electronic
characterization has only been successful relatively
recently\cite{NitzanetAl03Science,Tour00ACR}.  Two-terminal
configurations include nanopore\cite{RallsetAl88PRL}
structures\cite{ChenetAl99Science,ReedetAl01APL}, crossed wires
\cite{CollieretAl99Science,CollieretAl00Science,KushmericketAl02PRL},
mechanical break
junctions\cite{RuitenbeeketAl96RSI,ReedetAl97Science,ScheeretAl98Nature,ReichertetAl02PRL,SmitetAl02Nature},
and scanned probe approaches (scanning tunneling microscope
(STM)\cite{JoachimetAl95PRL,YazdanietAl96Science,DattaetAl97PRL,StipeetAl98Science,DonhauseretAl01Science,RamachandranetAl03Science}
or conducting probe atomic force microscope
(AFM)\cite{WoldetAl01JACS,CuietAl01Science}).

The resulting two-terminal devices exhibit a number of interesting phenomena,
including molecular rectification\cite{MetzgeretAl97JACS}, 
negative differential resistance (NDR)\cite{ChenetAl99Science}, and
switching between low and high conductance
states\cite{CollieretAl99Science,CollieretAl00Science,ReedetAl01APL,DonhauseretAl01Science,RamachandranetAl03Science}.
Several explanations for these
effects have been proposed, including: the electronic structure of the
molecular orbitals\cite{SeminarioetAl00JACS,SeminarioetAl01JPCA};
conformational changes in the
molecules\cite{SeminarioetAl98JACS,SeminarioetAl00JACS,EmberlyetAl01PRB,SolaketAl02ESSL,TroisietAl02JACS};
vibronic-assisted inelastic
tunneling\cite{GaudiosoetAl00PRL,KuznetsovetAl01JCP,TroisietAl03JCP};
and dynamic switching of contact
bonds\cite{RamachandranetAl03Science}.  A central question is whether
the nonlinear effects are intrinsic to the molecules themselves, or
the molecule-metal contact.  {\em Three-terminal} devices provide added
tunability to test candidate explanations for nonlinear conduction 
in these systems.

Three-terminal nanometer-scale single molecule transistors
(SMTs)\cite{ParketAl00Nature,ParketAl02Nature,LiangetAl02Nature,ParketAl03TSF,KubatkinetAl03Nature,YuetAl04NL,YuetAl03AC}
are a rich physical system, sensitive to molecular vibrational
modes\cite{ParketAl00Nature} and exhibiting correlated many-body
states (the Kondo
effect)\cite{ParketAl02Nature,LiangetAl02Nature,YuetAl04NL}.  Such
devices are tools for examining physics on the nanometer scale, and
promise to be the ultimate limit of electronic device
scaling.

In this paper we report two different nonlinear conductance phenomena
in molecule-scale three-terminal devices.  First, we show conduction
in C$_{60}$-based single-molecule transistors consistent with Kondo
physics.  The data indicate that inelastic couplings to vibrational
resonances in the molecule lead to enhanced transport at finite bias
in the Kondo regime.  We also show that nanometer-scale metal
junctions can, under certain circumstances, exhibit pronounced
negative differential resistance tunable by a proximal gate
electrode.  We argue that this NDR
most probably originates from the adsorption of a specific contaminant
at the nanometer-scale interelectrode gap.

\section{Technique}

\begin{figure}[h!]
\begin{center}
\includegraphics[clip, width=8cm]{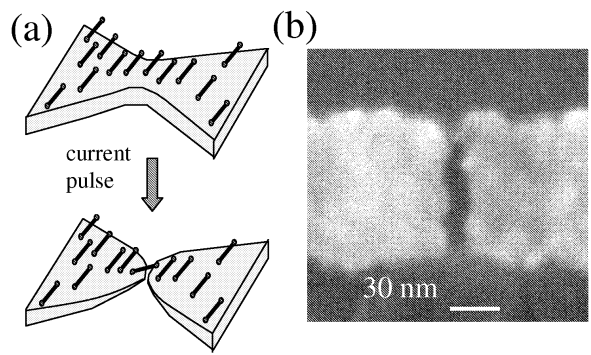}
\end{center}
\vspace{-5mm}
\caption{\label{fig:electromig}
(a) Diagram of electromigration technique for fabrication of closely spaced electrodes bridged by individual molecules.  (b) Micrograph of typical result, with 
electrodes separated by $<$~3~nm.}
\vspace{-5mm}
\end{figure}

Devices are prepared on an oxidized (200~nm), degenerately doped $p+$
Si substrate that is used as an underlying gate electrode.  The
fabrication technique, based on controlled electromigration of
lithographically fabricated metal constrictions\cite{ParketAl99APL},
is shown in Fig.~\ref{fig:electromig} and described in further detail
in Ref.~\cite{YuetAl04NL}.  A Ti (1.5~nm)/Au (15~nm) metal
constriction between two larger pads made by electron beam
lithography, e-beam evaporation, and lift-off.  The surface is then
exposed to oxygen plasma for 1~min. to remove any organic residue from
the lithography process.  Batches of as many as 60 constrictions are
fabricated on a single chip.  The transverse dimension of the
narrowest portion of the constriction is typically 80~nm.

For C$_{60}$-based devices, a dilute solution (1 mg C$_{60}$/ 4~mL
toluene) of molecules is spin cast (900~rpm) onto the array of
constrictions.  The wafer is then placed in a variable-temperature
vacuum probe station that is evacuated to $< \sim 10^{-5}$~mB by
turbopump.  Control devices are treated similarly, but without the
fullerenes.  Various common solvents have been examined, including
toluene, tetrahydrofuran (THF), and methyl chloride.
No particular precautions have been taken to avoid exposure to
ambient atmosphere during the spinning process.

At room temperature a semiconductor parameter analyzer is used to
sweep an applied voltage across the junction until the junction
conductance begins to degrade from electromigration.  This
``prebreaking'' is stopped once the junction resistance exceeds a few
hundred Ohms.  Subsequent examination of such junctions by scanning
electron microscope (SEM) indicates that the junctions are nearly
broken, so that the applied voltage drops almost entirely across the
now nano-scale constriction.

The sample area is then cooled to 4.2~K by flowing
liquid cryogens, while the chamber is
isolated from the turbo pumping system.  By carefully heating the
sample stage during this cooling process, stray gases are cryoadsorbed
to the chamber walls and a carbon felt ``'sorb'' rather than onto the
sample.  Cryopumping creates ultrahigh vacuum (UHV) conditions
at the sample while at low temperatures.
Further electromigration is performed at 4.2~K, breaking the
constrictions into separate source and drain electrodes while in UHV.

Figure~\ref{fig:electromig}b shows an SEM image of such an electrode
pair after warming to room temperature.  From all the constrictions
that begin intact at room temperature, measurable conduction at a
source-drain bias of 100~mV is found at the conclusion of the
electromigration procedure for 70\% of devices.  Because metal
surfaces anneal and reconstruct upon warming, it is difficult to
perform precise measurements of electrode morphology.  However, it
appears that electrodes must be spaced by 1-3~nm for measurable
conduction.

All measurements reported here were performed at dc using a
semiconductor parameter analyzer.  For each electrode pair, one
electrode (source) is defined as ground, and $I_{\rm D}$ is measured
as a function of $V_{\rm D}$ for various values of gate voltage,
$V_{\rm G}$.  Differential conductance, $dI_{\rm D}/dV_{\rm D}$, is
then computed numerically.  While numerical differentiation can be
noisy, this procedure has the benefit of working over a very broad
range of conductances.

In experiments done on the C$_{60}$-decorated junctions, the devices
are usually thermal cycled once over a period of two days without
venting the chamber.  In the NDR experiments discussed below,
electrodes were thermally cycled several times over three to five
days.  Device characteristics measured at low temperatures varied
after each thermal cycle, presumably due to electrode surface
reconstruction, molecule migration, and possible adsorption of
contaminants such as water.


\section{Device physics}
\subsection{Single-electron devices}

All successful nanometer-scale single-molecule transistors that have
been
demonstrated\cite{ParketAl00Nature,ParketAl02Nature,LiangetAl02Nature,ParketAl03TSF,KubatkinetAl03Nature,YuetAl04NL,YuetAl03AC}
act as single-electron transistors (SETs)\cite{GrabertetAl92book}.  A
SET consists of an island coupled by tunnel barriers ($R > R_{\rm
K}\equiv h/e^{2}$) to source and drain electrodes, with additional
capacitive coupling to a gate electrode.  Accounting for
electron-electron interactions on the island via a classical
capacitance, one may define an electron addition energy, $E_{\rm c}$,
associated with changing the charge of the island by a single
electron.  For $k_{\rm B}T << E_{\rm c}$, it is energetically
forbidden at zero source-drain bias to change the charge state of the
island.  The resulting suppression of transport is Coulomb
blockade.  Higher order processes ({\it e.g.} cotunneling) can allow
transport in the classically blockaded regime.

\begin{figure}[h!]
\begin{center}
\includegraphics[clip, width=8cm]{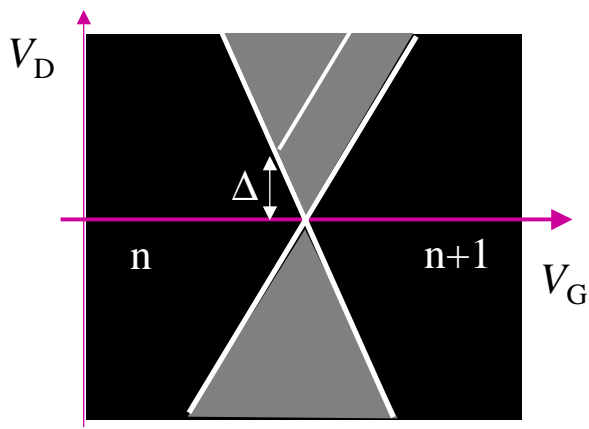}
\end{center}
\vspace{-5mm}
\caption{\label{fig:coulblock}
(a) Schematic map of differential source-drain conductance of asingle-electron transistor as a function of bias and gate voltage.  Brightness scale runs from zero conductance (black) to high conductance (white).  Arrow indicates a resonance due to tunneling into an excited state with energy $\Delta$ above the lowest unoccupied island state.}
\vspace{-5mm}
\end{figure}

It is possible to use $V_{\rm G}$ to lift the blockade at zero
source-drain bias by adjusting the potential of the island so that it
is energetically degenerate to change the charge of the island by one
electron.  Similarly, current may flow once $e V_{\rm
D}$ exceeds the energy addition threshhold.  Mapping differential
conductance as a function of $V_{\rm D}$ and $V_{\rm G}$, one finds a
result like that shown in Fig.~\ref{fig:coulblock}: diamond-shaped
blockaded regions where the charge on the island is quantized,
separated by a ``crossing point'' where, at $V_{\rm D} = 0$ and with
increasing $V_{\rm G}$, the occupation of the island changes from $n$
to $n+1$ electrons.  The lower the capacitive coupling to the gate
relative to the source and drain couplings, the more 
change in $V_{\rm G}$ necessary to change the charge state of the
island.

For a small island $E_{\rm c}$ includes both
the Coulomb repulsion, $U_{i}$, and the spacing between the $i$th and
$i+1$th single-particle electronic states, $\Delta_{i}$.  When
inelastic processes are allowed, at sufficient bias an electron may
tunnel from the source into {\em excited} single-particle states of
the island, as well as the lowest available single-particle
state.  Transport at high bias is enhanced by these additional
channels, manifested as peaks in the differential conductance that
parallel the edges of the blockaded diamonds.

When the island is a nanoscale molecule, several issues arise.  First,
source and drain couplings are established by overlap of the molecular
wavefunction with that of the conduction electrons of the leads.
These couplings broaden the energies of the single-particle states by
$\Gamma_{\rm S}, \Gamma_{\rm D}$, and depend crucially on the details
of the metal surface, the molecular orientation, and the chemical
nature of the molecular binding.  Furthermore, the classical
capacitance is so small that electrostatic charging energies in excess
of 100~meV are reasonable.  Similarly, the spacings between the
single-particle molecular states can easily be hundreds of meV or
more.  While there is some evidence that significant renormalizations
of both these quantities may occur\cite{KubatkinetAl03Nature}, most
experiments\cite{ParketAl00Nature,ParketAl02Nature,LiangetAl02Nature,ParketAl03TSF,YuetAl04NL}
find accessing more than two charge
states of a single molecule to be extremely challenging.

Because of these large level spacings, tunneling through excited
electronic states is generally not seen in SMTs.  However, molecules
do exhibit vibrational excitations at the meV scale.  Observations of
tunneling through such excited states have been
reported\cite{ParketAl00Nature,ParketAl02Nature,LiangetAl02Nature,ParketAl03TSF,YuetAl04NL},
as have vibrational effects in other gated molecular
devices\cite{ZhitenevetAl02PRL}.

A single-electron device with an unpaired spin may demonstrate
nontrivial transport
physics\cite{GGetAl98Nature,CronenwettetAl98Science} due to the Kondo
effect\cite{Kondo64PTP}, the formation of a many-body state comprising
the unpaired spin and an antiferromagnetically coupled screening cloud
of conduction electrons in the leads.  The energy scale, $k_{\rm
B}T_{\rm K}$, for the Kondo effect depends exponentially on the total
coupling $\Gamma = \Gamma_{\rm S}+\Gamma_{\rm D}$.  It is possible to
have a high Kondo temperature while still having a low total
conductance, since the conductance is limited by the smaller
of $\Gamma_{\rm D},\Gamma_{\rm S}$ while $T_{\rm K}$ is set by
their total. 

The signature of a Kondo resonance in a single-electron device is the
appearance of a peak in the differential conductance near zero
source-drain bias when the number of charges on the island changes
from even to odd.  In a well-coupled system for $T << T_{\rm K}$ the
peak conductance on resonance is expected to saturate to $2e^{2}/h$.
This value may be significantly reduced if the source and drain
couplings are very asymmetric.  For $T >\sim T_{\rm K}$, the peak
conductance is expected to decrease logarithmically with temperature.
The shape of this resonance ($I_{\rm D}$ as a function of $V_{\rm D}$)
may deviate from a pure Lorentzian due to Fano-type competition
between Kondo resonant transport and other
channels\cite{MadhavanetAl98Science,MadhavanetAl01PRB,GoresetAl00PRB}.
In an external magnetic field, $B$, large enough to Zeeman split the
localized, unpaired spin, the Kondo peak is expected to split by an
amount $g \mu_{\rm B} B$, where $\mu_{\rm B}$ is the Bohr magneton and
$g$ is the Land{\'e} factor.  Kondo physics has been reported in molecular
devices incorporating metal
ions\cite{ParketAl02Nature,LiangetAl02Nature}.  Our observation of
C$_{60}$-based devices is described below.  We discuss
vibrational resonances in the Kondo regime, a phenomenon only
observable in molecular devices.

\subsection{Negative differential resistance}

Negative differential resistance is well known and understood in
two-terminal devices such as $p-n$ junctions\cite{Esaki58PR},
quantum well resonant tunneling diodes\cite{SollneretAl83APL}, and
scanning tunneling spectroscopy (STS)
measurements\cite{LyoetAl89Science}.  The basic mechanism for each
involves energy-specific conduction of charge carriers; as the
source-drain bias is swept into and out of the relevant energy range,
the current varies nonmonotonically, leading to NDR over some range of
bias voltage.  In Esaki diodes the current peaks due to the alignment
of the valence and conduction bands across the $p-n$ junction over a
limited range of bias.  In quantum well devices the increase in
conduction occurs when the energy of the injected carriers is resonant
with a quasi-bound state of the well.  Here we will explore the origin
of NDR in the STS measurements\cite{XueetAl99PRB} in more detail
because we believe a similar mechanism is responsible for our NDR
observations.

Consider a tunneling spectroscopy measurement with electrons
flowing from a probe tip to a conducting substrate,
with a molecule attached to the substrate.
To explain the observation of NDR 
one must abandon the common notion of a featureless tip density
of states, and describe the current through the STS system with the
general expression:
\begin{equation}
I = \int_{\mu_{\rm S}}^{\mu_{\rm P}} T(E,V)dE,
\end{equation}
where $\mu_{\rm S}$ and $\mu_{\rm P}$ are the electrochemical
potential of the metallic substrate and probe tip, respectively, with
$\mu_{\rm S}=\mu_{\rm P}+eV$.  Here $T(E,V)$ is the tunneling
transmission of the system.  Using the transfer Hamiltonian formalism,
one can relate $T(E,V)$ to the product of the local densities of
states (LDOS) $\rho_{\rm mol}(E)$ and $\rho_{\rm tip}(E)$ of the molecule
and the tip.  The current through the STS system is proportional to
the convolution of $\rho_{\rm mol}$ and $\rho_{\rm tip}$ within the
energy range defined by $\mu_{\rm S}$ and $\mu_{\rm P}$.  

\begin{figure}[h!]
\begin{center}
\includegraphics[clip, width=8cm]{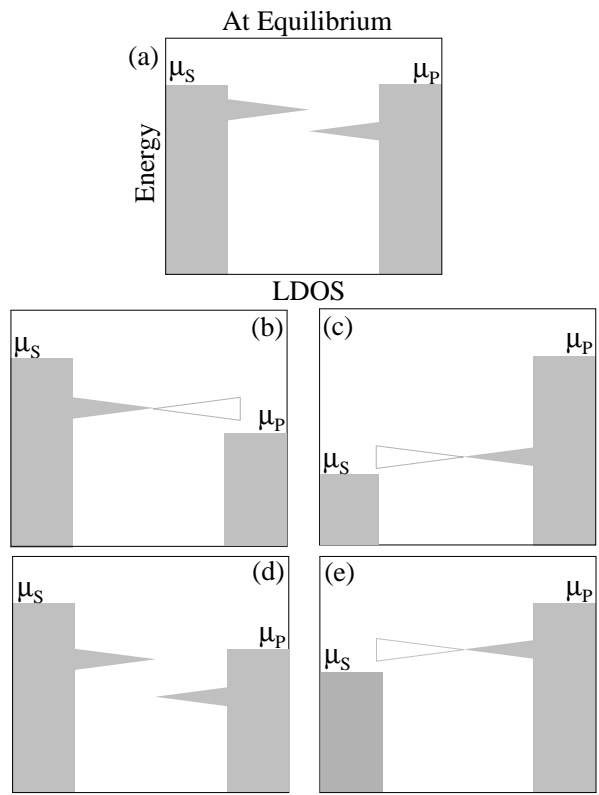}
\end{center}
\vspace{-5mm}
\caption{\label{fig:NDRexp}
Energetics of substrate/molecule and tip apex/probe in STS
measurements on molecules, after \protect{\cite{XueetAl99PRB}}. Sharp
features represent local densities of states of the molecule (left)
and the tip apex (right).  Weak coupling allows a potential difference
to develop between the molecule and substrate under bias. (a) At
equilibrium, with substrate chemical potential, $\mu_{\rm
S}$, and probe chemical potential, $\mu_{\rm P}$; (b) A positive bias $V$
is applied to the probe, and weak apex-probe coupling is assumed,
leading to the apex potential $V_{\rm Apex} < V$.  The LDOS of the
apex atom moves up relative to $\mu_{\rm P}$ under bias, and NDR
results; (c) A negative bias is applied to the probe, again with
$|V_{\rm Apex}| < |V|$.  The LDOS of the apex atom move down relative
to $\mu_{\rm P}$, and NDR results; (d) A positive bias is applied to
the probe, with strong apex-probe coupling that forces $V_{\rm Apex} =
V$.  No NDR results because of relative positions of molecule and apex
LDOS (e) A negative bias is applied to the probe, with $V_{\rm Apex} =
V$.  Since the LDOS of the molecule must move down with respect to the
$\mu_{\rm P}$ under bias, NDR results.}
\vspace{-5mm}
\end{figure}

Suppose both the $\rho_{\rm mol}$ and the $\rho_{\rm tip}$ contain
sharp peaks (such as those caused by localized surface states, for
example\cite{Lang97PRB}).  As the bias is applied the LDOS shift
relative to one another, and it is possible for the energies of the
sharp features to become commensurate.  If this alignment happens when
one sharp feature is occupied and the other is empty, one may observe NDR.  
This situation is illustrated in Figure~\ref{fig:NDRexp}b,c,e.
As the LDOS peaks move toward (away from)
each other in energy while the bias is swept, the current will
increase (decrease).

It is important to consider whether one expects to find NDR under both
signs of bias within this picture; both cases are observed in STS.  As
explained in Ref.~\cite{XueetAl99PRB}, the critical issue is whether
there is significant voltage dropped between the apex of the probe tip
(source of the sharp LDOS peak) and the bulk of the probe tip (and
similarly between the molecule and the substrate).  If the apex and
bulk of the tip are essentially at the same chemical potential, then
NDR is only expected under one bias
direction (Figure~\ref{fig:NDRexp}d,e).  Note that a weak coupling
between the apex and bulk of the tip, a necessary condition for the
observation of NDR under both signs of bias, would also imply poor
screening near the tip apex.  This is relevant when considering the
effects of a proximal gate electrode.

\section{Device characterization and statistics}

At present no microscopy technique allows
direct assessment of the presence of a single nanometer-scale molecule
at the source-drain interelectrode gap.  Therefore one must
infer such information from transport measurements and control
experiments.  Every electrode pair produced by the electromigration
process is different at the nanometer scale.  Similarly, the
positions, orientations, couplings to source, drain, and gate
electrodes, of possible molecules near the interelectrode gaps are all
probabilistic.  With this fabrication technique {\it a statistical
approach with large numbers of devices is essential}.

Conduction characteristics at low temperatures after electromigration
may be divided into four general classes:
\begin{itemize}
\item[A.] {\it No detectable source-drain current}.  The most likely
explanation, confirmed by SEM observation, is that the resulting
interelectrode gap is too large for detectable source-drain
conduction by either tunneling, thermionic, or field emission.

\item[B.] {\it Weak nonlinearity, no gate response}.  This is the
result most commonly seen in bare metal control devices, and 
presumably corresponds to electrodes sufficiently close (1-2~nm)
for some conduction, but without a molecule in the gap.

\item[C.] {\it Strong nonlinearity, no gate response}.  The
most probable explanation for such $I_{\rm D}-V_{\rm D}$ 
characteristics is that a molecule or metal nanoparticle is
positioned between the source and drain, but device geometry
results in extremely poor gate coupling.  

\item[D.] {\it Strong nonlinearity, gate response}.  These are the
devices of interest, and consist of either a molecule, group of
molecules, contaminant, or metal nanoparticle between
nanometer-separated source and drain.
\end{itemize}
Table~\ref{tab:stats} shows the statistics of all 724 devices
prepared using C$_{60}$ molecules.  

\begin{table}[h!]
\caption{\label{tab:stats}
Device yield statistics of C$_{60}$ SMTs.}
\begin{center}
\begin{tabular}{|c||c|c|}
\hline
$I_{\rm D}-V_{\rm SD}$ curve type & Interpretation & \% of devices \\
\hline
none & gap too large & 30\%\\
weakly nonlinear & no molecule & 38\% \\
strongly nonlinear, no gating & no gate coupling & 19\% \\
strongly nonlinear + gating & candidate SMT & 13\%\\
\hline
\end{tabular}
\end{center}
\end{table}

Given a gateable, nonlinear $I_{\rm D}-V_{\rm D}$ characteristic
consistent with Coulomb blockade, it is {\it essential} to consider
four other issues to help determine whether the device is a SMT or a
result of metal nanoparticles produced in the electromigration
process.  First, is the electron addition energy (the maximum
source-drain bias of the blockaded region) of a sensible size for the
molecule under consideration?  As mentioned above, the {\it minimum}
addition energy, even in the absence of molecular level spacings,
should be sensibly large for Coulomb charging of a nanoscale object in
a dielectric environment.  Addition energies significantly less than
50~meV should be considered carefully.

Second, are the number of accessible charge states of the device
reasonable for a molecule?  Solution-based electrochemistry provides
an upper limit to the number of valence states that one should
reasonably be able to explore.  In electrochemical experiments, the
excess molecular charge is compensated by the presence of ions in the
solvent that can be within a nanometer of the molecule.  This
screening should be more efficient than any possible in a conventional
field effect geometry, even with an extremely thin gate dielectric.
For example, C$_{60}$ may be reduced in solution only a few times.
Therefore if one finds that it is possible to add ten or twenty
electrons to a candidate device via gating ({\it i.e.} there are many
Coulomb blockade crossing points), it is extremely unlikely that the
active region of the candidate device is a single C$_{60}$ molecule.
This issue makes interpretation of some data\cite{YuetAl03AC}
challenging.

Third, are the particular gateable, nonlinear characteristics observed
only when molecules are present?  We have observed clear Coulomb
blockade with various electron addition energies in a small percentage
of bare metal control devices.  Clearly Coulomb blockade
characteristics alone are not sufficient to confirm
the molecular character of a SMT.  Similarly, the NDR effects
described below appear predominantly in devices prepared {\it without}
C$_{60}$ molecules, and therefore cannot represent a property of
C$_{60}$-based conduction.

Finally, are there features in the data that uniquely specify the
molecule?  Known vibrational resonances in particular can be extremely
useful, and have been identified in several
experiments\cite{ParketAl00Nature,ParketAl03TSF,YuetAl04NL}.

\section{Results and discussion}

\subsection{C$_{60}$ devices and Kondo physics}

\begin{figure}[h!]
\begin{center}
\includegraphics[clip, width=8cm]{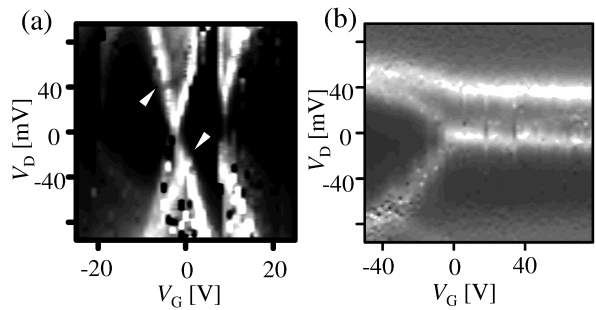}
\end{center}
\vspace{-5mm}
\caption{\label{fig:c60}
(a) Map of differential source-drain conductance of a C$_{60}$-based SMT at 5~K, as a function of bias and gate voltage.  Brightness scale runs from zero conductance (black) to $3 \times 10^{-6}$~S (white).  Arrows indicate 35~meV resonances due to vibrational mode of C$_{60}$. (b) Analogous map for a device exhibiting conduction consistent with Kondo physics.  Conductance range is from zero (black) to $1.5 \times 10^{-5}$~S (white).}
\vspace{-5mm}
\end{figure}

Figure~\ref{fig:c60}a is a conductance map of a C$_{60}$-based SMT
measured at 5~K.  Coulomb blockade is clearly evident, as is an uncontrolled change
in offset charge (the discontinuity at $V_{\rm G}=8$~V).
Note that the electron addition energy significantly exceeds 100~meV,
and that there are vibrational resonance features at $\sim$~35~meV, as
indicated by the arrows.  Much of the noise in the data are an
artifact of the numerical differentiation procedure.  Scans out to
higher gate voltages in both polarities revealed no further Coulomb
blockade crossing points.  This indicates that we were only able to
change the charge state of this device by a single electron.

Figure~\ref{fig:c60}b is a conductance map
for a SMT that appears consistent with Kondo physics. At gate
voltages less than -10~V, conductance is consistent with Coulomb
blockade, including a likely vibrational resonance at $\sim$~35~meV.
When $V_{\rm G}$ is swept past the charge degeneracy point, a
pronounced peak appears at zero bias.  Notice that the 35~meV peak
continues into this regime.  In fact both the 35~meV conductance and
the zero-bias conductance are enhanced relative to the conductances
measured in the Coulomb blockade regime.  Some indications of
vibrational signatures in the Kondo regime were present in an earlier
experiment\cite{LiangetAl02Nature}.  In the C$_{60}$ case shown here,
however, the vibrational mode is known.  We also note that enhanced
coupling to vibrational modes in the presence of a strong coupling
between adsorbed molecular charge and metal conduction electrons is
known to occur in other cases, such as the ``chemical enhancement''
contribution to surface-enhanced Raman
scattering\cite{MoskovitsetAl02TAP}.

\begin{figure}[h!]
\begin{center}
\includegraphics[clip, width=8cm]{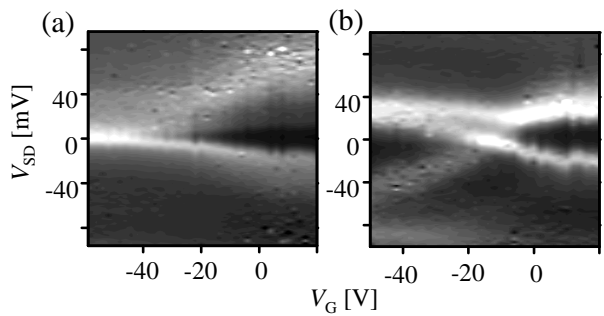}
\end{center}
\vspace{-5mm}
\caption{\label{fig:c60xition}
\small Example of an irreversible change from Kondo-like conduction (a) to conventional Coulomb blockade conduction (b) in a single device, likely
due to an uncontrolled rearrangement of the molecule-electrode tip geometry.
Conductance is from zero (black) to $1.8 \times 10^{-5}$~S (white).
}
\vspace{-5mm}
\end{figure}

A detailed analysis of several samples similar to that in
Fig.~\ref{fig:c60}b is presented in Ref.~\cite{YuetAl04NL}.  Several
points are noteworthy.  First, only $\sim$1-2\% of C$_{60}$-decorated
electrodes eventually resulted in devices that showed this Kondo-like
conductance map.  This is unsurprising when one recalls that $T_{\rm
K}$ depends exponentially on $\Gamma$, which itself is exponentially
sensitive to molecule-metal geometry.  Similarly, presumably because
of these steep dependences, we found it extremely difficult to perform
characterization of such devices over a broad temperature range due to
irreversible changes in device characteristics.
Figure~\ref{fig:c60xition} shows an example of this instability.  At
left is the conductance map of the device showing the transition to
having the zero-bias resonance.  During an attempt to characterize the
temperature dependence of the zero-bias peak, the device spontaneously
changed to a different configuration.  Subsequent measurements of the
device yielded conductance characteristics like that shown in the plot
at right, looking like Coulomb blockade with poor gate coupling.  The
typical maximum conductance remains $\sim 10^{-5}$~S despite this
change in coupling, consistent with the idea that the couplings to the
leads are asymmetric, as discussed above.

From the limited $T$-dependence data acquired, the zero-bias peak is
essentially independent of $T$ below $\sim$~20~K.  Within the Kondo
picture the width of the peak is proportional to $T_{\rm K}$ when $T$
is sufficiently below $T_{\rm K}$.  Our data are consistent with
values of $T_{\rm K}$ of around 100~K or more.  This energy scale is
so large that resolving Zeeman splitting of the peak is unfeasible.

The high Kondo temperatures inferred in the C$_{60}$ system may
explain another observation\cite{KellyetAl96Science} in scanning
tunneling microscopy (STM) using probe tips made from C$_{60}$
adsorbed on platinum-iridium.  Those STM measurements indicated a
surprisingly narrow, peaked density of states for the tips.  One
explanation for such a feature would be a Kondo resonance between the
physisorbed C$_{60}$ and the tip, perhaps surviving even at 300~K.
Experiments on C$_{60}$ SMTs with Pt electrodes are planned.
We note that fabricating SMTs that are mechanically and electrically
stable and exhibit Kondo physics at room temperature would have
significant implications for practical devices.  Such SMTs would be
switchable by gating from a Coulomb blockaded state to resonant
conduction with conductances approaching $2e^2/h$, the theoretical
maximum per molecule.  Such devices would remain vulnerable to offset
charge problems characteristic of single-electron
architectures\cite{Likharev99IEEE}.  They would, however, offer the
only known means of modulating such high conductances per molecule
apart from larger structures involving perfectly contacted carbon
nanotubes.

\subsection{Negative differential resistance}

\begin{figure}[h!]
\begin{center}
\includegraphics[clip, width=8cm]{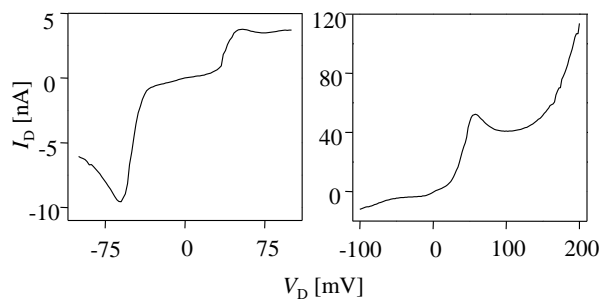}
\end{center}
\vspace{-5mm}
\caption{\label{fig:ndrIV}
\small Examples of $I_{\rm D}-V_{\rm D}$ characteristics ($V_{\rm G}=0$, $T = 4.2$~K) for control samples (no C$_{60}$) exposed to non-UHV 
conditions repeatedly.  NDR can occur for either one or both bias
polarities.}
\vspace{-5mm}
\end{figure}

In a series of control experiments, electromigrated control samples
that were exposed to various solvents (e.g. THF, toluene, methyl
chloride) prior to being place in the vacuum probe station were
thermally cycled multiple times over the course of three to five days.
In 10\% of 175 multiple thermal cycled control devices we observed
regions of NDR in their $I_{\rm D}-V_{\rm D}$ curves at 4.2~K, such as
those shown in Figure~\ref{fig:ndrIV}.

\begin{figure}[h!]
\begin{center}
\includegraphics[clip, width=8cm]{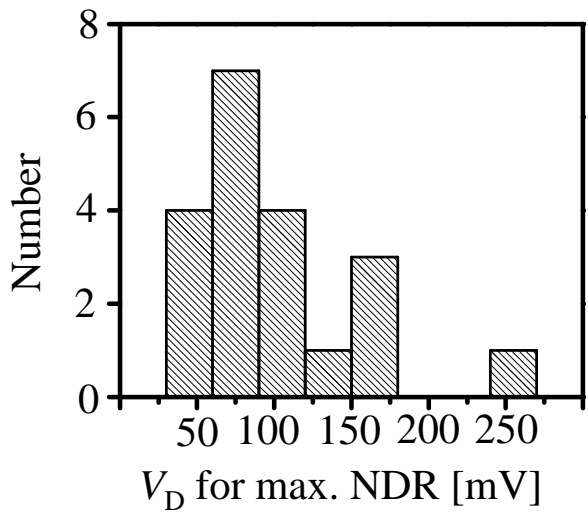}
\end{center}
\vspace{-5mm}
\caption{\label{fig:hist}
For $V_{\rm G}=0$, distribution of bias voltages at which
maximum NDR occurs.  The histogram is singly peaked, suggesting that a single
specific mechanism is producing the NDR effect. }
\vspace{-5mm}
\end{figure}

We suggest that the NDR results from sharp features in the densities
of states of the source and drain electrodes due to adsorbed
impurities.  The role of unintended adsorbates seems clear, since the
NDR tends to appear as surfaces are exposed for significant periods in
non-UHV conditions.  We believe that there is {\it one
particular kind of adsorbate} that causes the NDR in our control
devices.  The evidence for our assertion is shown in
Fig.~\ref{fig:hist}, a histogram of the bias position (at zero
gate voltage) where the maximum NDR occurs for these devices.  The
histogram is peaked prominently near 75~mV, suggesting that the NDR in
these devices has a well-defined, common origin, rather than being a
random phenomenon.

\begin{figure}[h!]
\begin{center}
\includegraphics[clip, width=8cm]{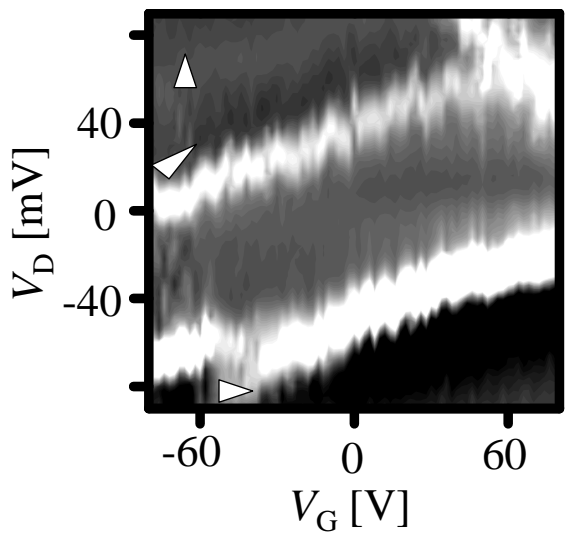}
\end{center}
\vspace{-5mm}
\caption{\label{fig:gatendr}
Differential conductance map of a control device
showing gateable NDR.  Brightness runs from $-8 \times 10^{-8}$~S
(black) to $2 \times 10^{-7}$~S (white).  Arrows indicate NDR
regions.}
\vspace{-5mm}
\end{figure}

Similar to STS experiments where NDR were
observed\cite{LyoetAl89Science,XueetAl99PRB}, we also observe in these
devices that NDR can appear in either only one or both bias
directions.  About 30\% of the NDR devices exhibit NDR in both bias
polarities.  All devices that exhibit NDR in both bias directions are
also gateable.  The conductance map of such a device is shown in
Figure~\ref{fig:gatendr}.  Regions of NDR are indicated by the arrows.

The concurrency of NDR in both bias directions and the
gateablity of the devices reflects the extended screening length in
the metal-molecule junction when the apex of the electrode is
weakly coupled to the bulk of the electrode.  If the screening length
is long enough for the emergence of a voltage drop between the apex
and the bulk of the electrode, then it would likely to be long
enough for the gate to influence the electrostatic potential of the
metal-molecule junction.

The fact that the nonlinear conduction can be modulated by $V_{\rm G}$
even in the absence of Coulomb blockade strongly constrains
alternative NDR explanations.  For example, vibrationally mediated NDR
sometimes observed in STS experiments on adsorbed
molecules\cite{GaudiosoetAl00PRL} is only expected to occur at biases
that correspond to specific mechanical modes of the adsorbate.  Since
those vibrational modes are determined by molecular structure rather
than electrostatics, gate-independent NDR would be expected from this
mechanism.  Similarly, the lack of correlation of NDR with Coulomb
charging implies that the molecular orbital effects
suggested\cite{SeminarioetAl00JACS} to explain some nanopore NDR
observations\cite{ChenetAl99Science} are not relevant.

Experiments are ongoing to determine the particular adsorbates
responsible for the NDR.  As with the Kondo physics discussed above, a
reliable method to create stable, nanometer-scale NDR devices with
significant peak-to-valley ratios would have significant practical
implications.

\section{Summary}

We have employed an electromigration technique to create three
terminal devices with source and drain electrodes separated on the
single nanometer scale.  With these devices we have demonstrated
single-molecule transistors based on C$_{60}$ that operate as
single-electron devices that can access two molecular charge states.
Some of these C$_{60}$ SMTs exhibit conductance properties consistent
with Kondo physics.  Kondo transport here is indicated by the presence
of a zero-bias peak in the conductance for one SMT charge state, and
Coulomb blockade for the other.  The role of C$_{60}$ is indicated by
resonance features in the conductance energetically compatible with
known molecular vibrational modes.  Because of the resonant nature of
conduction in the Kondo regime, this physics holds out the promise of
SMTs with a conductance per molecule in the ``on'' state approaching
the theoretical maximum, $2e^2/h$.

Control devices left exposed to non-UHV conditions 
demonstrate a propensity to exhibit negative differential resistance.
We have shown that this NDR may be modulated via $V_{\rm G}$, and
propose that its explanation lies in sharp peaks in the local
densities of states of the source and drain electrodes due to adsorbed
impurities.  We argue that the peaked distribution of NDR voltage
characteristics implies that a single type of impurity is responsible
for this nonlinear conduction.  An improved understanding of this
system may allow the controlled engineering of molecule-scale NDR
devices.

\ack

The authors thank K. Kelly and A. Osgood for
STM characterization, and P.L. McEuen, P. Nordlander, H. Park,
D.C. Ralph, A. Rimberg, J.M. Tour, and R.L. Willett for useful
conversations.

This work was supported by the Robert A. Welch Foundation, the
Research Corporation, the David and Lucille Packard Foundation, and the National
Science Foundation.


\end{document}